\documentclass[9pt,twocolumn,twoside]{pnas-new}

\usepackage[load-configurations=abbreviations,range-phrase=--,range-units=single,detect-all]{siunitx}
\usepackage{wasysym}
\usepackage[percent]{overpic}
\usepackage{marvosym}

\definecolor{mustard}{RGB}{204,204,0}
\definecolor{limegreen}{RGB}{0,204,103}
\definecolor{lightGrey}{rgb}{0.6,0.6,0.6}

\DeclareSIUnit\rpm{rpm}

\templatetype{pnasresearcharticle} 

\title{\Huge Conching chocolate is a prototypical transition from frictionally jammed solid to flowable suspension with maximal solid content}

\author[a,1]{Elena Blanco}
\author[a,1,*]{Daniel J. M. Hodgson}
\author[a,b,1]{Michiel Hermes}
\author[a,c]{Rut Besseling}
\author[d,e]{Gary L. Hunter}
\author[d]{Paul M. Chaikin}
\author[a,f]{Michael E. Cates}
\author[g]{Isabella Van Damme}
\author[a]{Wilson C. K. Poon}

\affil[a]{School of Physics and Astronomy, The University of Edinburgh, King's Buildings, Peter Guthrie Tait Road, Edinburgh EH9 3FD, UK}
\affil[b]{Soft Condensed Matter, Debye Institute for Nanomaterials Science, Utrecht University, Princetonplein 5, 3584 CC Utrecht, The Netherlands}
\affil[c]{InProcess-LSP, Kloosterstraat 9, 5349 AB Oss, Netherlands}
\affil[d]{Center for Soft Matter Research (CSMR), Department of Physics, New York University}
\affil[e]{ExxonMobil Research and Engineering Corporate Strategic Research, 1545 Route 22 East, Annandale, New Jersey 08801}
\affil[f]{DAMTP, Centre for Mathematical Sciences, Wilberforce Road, Cambridge, CB3 0WA, UK}
\affil[g]{Mars Chocolate UK Ltd., Dundee Road, Slough, SL1 4JX, UK}
\leadauthor{Blanco}

\significancestatement{Chocolate conching is the process in which an inhomogeneous mixture of fat, sugar and cocoa solids is transformed into a homogeneous flowing liquid.  Despite the popularity of chocolate and the antiquity of the process, until now there has been poor understanding of the physical mechanisms involved.  Here, for the first time, we show that two of the main roles of conching are the mechanical breakdown of aggregates and the reduction of inter-particle friction through the addition of a dispersant. Intriguingly, the underlying physics we describe is related to the popular stunt of `running on cornstarch'.}

\authorcontributions{Author contributions: WP, PC, MH, MC and IVD conceptualised the work. EB, DH, GH and RB performed experiments.  EB, DH, MH, RB  and GH analysed and interpreted data. DH and WP prepared the manuscript.}
\authordeclaration{Conflict of interest statement: This work was funded by Mars Chocolate UK Ltd..}
\equalauthors{\textsuperscript{1}EB, DH and MH contributed equally to this work.}
\correspondingauthor{\textsuperscript{*}To whom correspondence should be addressed. E-mail: danieljmhodgson@gmail.com}


\begin{abstract} 
The mixing of a powder of 10-\SI{50}{\micro\meter} primary particles into a liquid to form a dispersion with the highest possible solid content is a common industrial operation. Building on recent advances in the rheology of such `granular dispersions', we study a paradigmatic example of such powder incorporation: the conching of chocolate, in which a homogeneous, flowing suspension is prepared from an inhomogeneous mixture of particulates, triglyceride oil and dispersants. Studying the rheology of a simplified formulation, we find that the input of mechanical energy and staged addition of surfactants combine to effect a considerable shift in the jamming volume fraction of the system, thus increasing the maximum flowable solid content. We discuss the possible microscopic origins of this shift, and suggest that chocolate conching exemplifies a ubiquitous class of powder-liquid mixing.
\end{abstract}

\dates{This manuscript was compiled on \today}
\doi{\url{www.pnas.org/cgi/doi/10.1073/pnas.XXXXXXXXXX}}

\begin{document}

\verticaladjustment{-2pt}

\maketitle
\thispagestyle{firststyle}
\ifthenelse{\boolean{shortarticle}}{\ifthenelse{\boolean{singlecolumn}}{\abscontentformatted}{\abscontent}}{}


\dropcap{T}he incorporation of liquid into dry powder with primary particle size in the granular range ($\sim 10$-\SI{50}{\micro\meter}) to form a flowing suspension with solid volume fraction $\phi \gtrsim 50\%$ is important in many industries \cite{Kaye1997}. Often, maximising solid content is a key goal. Cements for building or bone replacement and ceramic `green bodies' are important examples, where higher $\phi$ improves material strength \cite{Green2004}. Another example is chocolate manufacturing, where high solid content (= lower fat \cite{Tao2016}) is achieved by `conching'.

Conching \cite{Gutierrez2017}, invented by Rodolphe Lindt in 1879, is important for flavor development, but its major physical function is to turn an inhomogeneous mixture of particulates (including sugar, milk solids and cocoa solids) and cocoa butter (a triglyceride mixture) into a homogeneous, flowing suspension (liquid chocolate) by prolonged mechanical action and the staged addition of dispersants. In this paper, we focus on this effect, and seek to understand how mechanical action and dispersants together transform a non-flowing, inhomogeneous mixture into a flowing suspension, a process that has analogs in, e.g., the ceramics and pharmaceuticals sectors \cite{Kaye1997}.

We find that the key physical processes are friction-dominated flow and jamming.  Specifically, two of the key rheological parameters in chocolate manufacturing, the yield stress, $\sigma_\textrm{y}$, and the high-shear viscosity, $\eta_\textrm{2}$, are controlled by how far the volume fraction of solids, $\phi$, of the chocolate formulation is situated from the jamming volume fraction, $\phi_{\textrm{J}}$.  We demonstrate that the first part of the conche breaks apart particulate aggregates, thus increasing $\phi_{\rm J}$ relative to the fixed mass fraction. In the second part of the conche, the addition of a small amount of dispersant reduces the inter-particle friction and further raises $\phi_{\rm J}$, in turn reducing $\sigma_{\rm y}$ and $\eta_2$, resulting in fluidization of the suspension, i.e. a solid to liquid transition. Such `$\phi_{\rm J}$ engineering' is common to diverse industries that rely on the production of high-solid-content dispersions.

\section*{Shear thickening suspensions}
\begin{figure}[t]
\includegraphics[width=0.86\columnwidth]{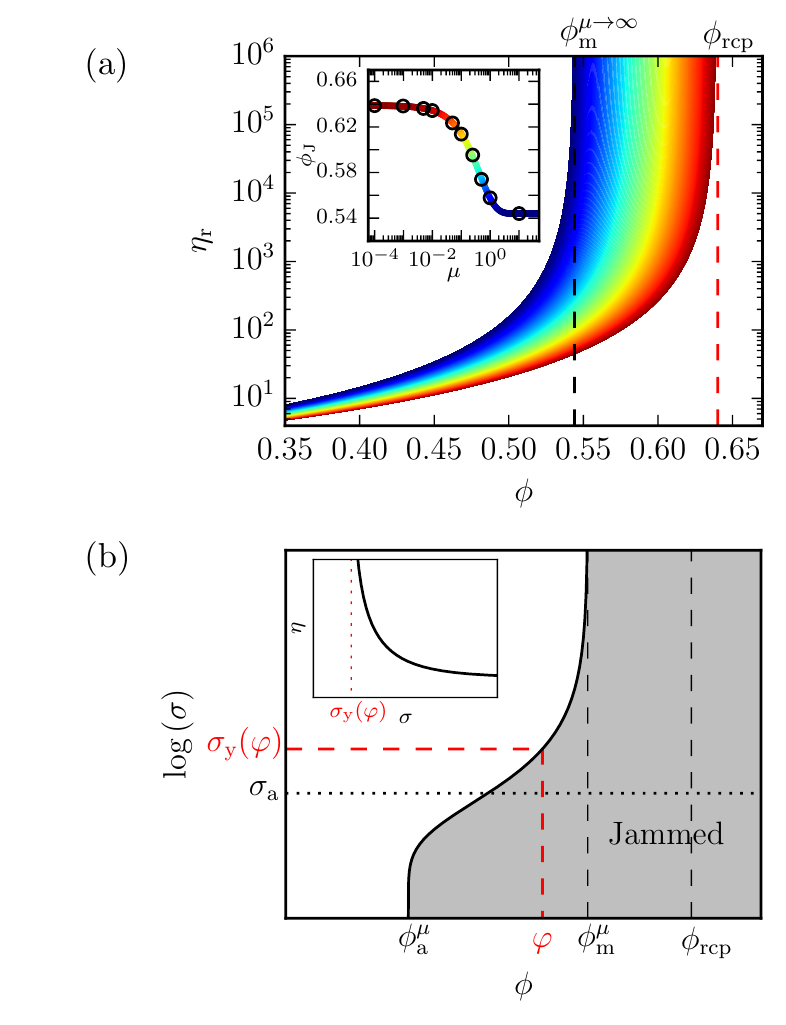}
\caption{(a) The high-shear viscosity of suspensions of granular hard spheres normalised by the solvent viscosity, $\eta_{\textrm{r}}$, plotted against the volume fraction $\phi$, with friction coefficient increasing from $\mu = 0$ (red), diverging at $\phi_{\textrm{rcp}}$, to $\mu \to \infty$ (blue), diverging at $\phi_{\textrm{m}}^{\mu \to \infty}$. Inset: The jamming volume fraction, $\phi_{\textrm{J}}$, where $\eta_{\textrm{r}}$ diverges, as a function of the coefficient of static friction $\mu$ (replotted from \cite{silbert2010jamming}). (b) The jamming state diagram of a frictional granular suspension with inter-particle adhesion. The adhesive strength is set by $\sigma_{\textrm{a}} \gg \sigma^\star$. Shaded region = jammed. The flow curve of a suspension with volume fraction $\varphi$ is shown in the inset: it has a yield stress $\sigma_{\textrm{y}}(\varphi)$.}
  \label{fig:FrictionDependentPhiJ}
\end{figure}

We first review briefly recent advances in granular suspension rheology \cite{Seto2013,WC1,WC2,mari2014shear,Guy2015,Lin2015,Royer2016,Hermes2016,Colin2017,clavaud2017revealing}.
The viscosity of a high-$\phi$ granular suspension increases from a low-stress Newtonian value when the applied stress, $\sigma$, exceeds some onset stress, $\sigma^\star$, reaching a higher Newtonian plateau at $\sigma \gg \sigma^\star$: the suspension shear thickens. The low- and high-stress viscosities, $\eta_1$ and $\eta_2$, diverge as
\begin{equation}\label{eq:krieger-dougherty}
  \eta_{\textrm{r}} = A \left( 1 - \frac{\phi}{\phi^\mu_{\textrm{J}}(\sigma)} \right)^{-\lambda},
\end{equation}
where $\eta_{\textrm{r}} = \eta_{1,2}/\eta_0$ with $\eta_0$ the solvent viscosity, $A\simeq 1$, and $\lambda \simeq 2$ for spheres \cite{Maron1956,KD1959}.
The jamming point, $\phi_{\textrm{J}}$, is a function of both the inter-particle friction coefficient, $\mu$, and the applied stress, $\sigma$. The latter begins to press particles into contact when it exceeds $\sigma^\star$. With $\mu \to 0$, no shear thickening is observed, and $\eta_{\textrm{r}}$ diverges at random close packing, $\phi_{\textrm{J}} = \phi_{\textrm{rcp}}$. At finite $\mu$, the low-stress viscosity $\eta_1(\phi)$ still diverges at $\phi_{\textrm{rcp}}$, but $\eta_2(\phi)$, the high-stress viscosity, now diverges at some $\phi_{\textrm{J}} = \phi_{\textrm{m}}^\mu < \phi_{\textrm{rcp}}$. For monodisperse hard spheres, \autoref{fig:FrictionDependentPhiJ}(a), $\phi_{\textrm{rcp}} \approx 0.64$ and $\phi_{\textrm{m}}^{\mu \to \infty} \approx 0.54$ (where `$\infty$' in practice means $\mu \gtrsim 1$) \cite{silbert2010jamming,mari2014shear}. (Below we drop the `$\mu$' in $\phi_{\textrm{m}}^\mu$ unless it is needed.)
A granular suspension at $\phi > \phi_{\textrm{m}}$ cannot flow at high stress either steadily or homogeneously \cite{Hermes2016}: it shear-jams \cite{WC2}. Instead, theory \cite{WC2} and experiments \cite{Hodgson2015} suggest that it granulates.

The onset stress, $\sigma^\star$, correlates with the force to overcome an inter-particle repulsive barrier; typically $\sigma^* \sim d^{-\nu}$ with $\nu \lesssim 2$, where $d$ is the particle diameter \cite{Guy2015}.  For granular suspensions, $\sigma^{\star}$ is far below stresses encountered in liquid-powder mixing processes, so that they always flow with viscosity $\eta_2(\phi,\mu)$, which diverges at $\phi_\textrm{m} < \phi_\textrm{rcp}$. To formulate a flowable granular suspension with maximum solid content is therefore a matter of maximising $\phi_\textrm{m}$, e.g., by lowering $\mu$, \autoref{fig:FrictionDependentPhiJ}(a) (inset).

Inter-particle adhesion introduces another stress scale,  $\sigma_{\textrm{a}}$, characterising the strength of adhesive interactions \cite{guy2018constraint}. A yield stress, $\sigma_{\textrm{y}}$, emerges above some $\phi_{\textrm{a}}^\mu < \phi_{\textrm{m}}^\mu$ that is dependent on both adhesion and friction \cite{liu2017,guy2018constraint} (hence the $\mu$ superscript, which, again, we will drop unless needed), and diverges at $\phi_{\textrm{m}}^\mu$.

Competition between friction and adhesion gives rise to a range of rheologies \cite{guy2018constraint}. If $\sigma^{\star}/\sigma_{\textrm{a}}\ll 1$, the suspension shear thins at $\sigma > \sigma_{\textrm{y}}$ to the frictional viscosity, $\eta_2$. The state diagram such a system is shown schematically in \autoref{fig:FrictionDependentPhiJ}(b); the inset shows a typical flow curve. However, a suspension with $\sigma^{\star}/\sigma_{\textrm{a}}\gg1$ first shear thins at $\sigma > \sigma_{\textrm{y}}$, and then shear thickens as $\sigma$ exceeds $\sigma^\star$. Modifying system additives (e.g. removing polymeric depletants or adding surfactants) can increase $\sigma^{\star}/\sigma_{\textrm{a}}$ and change the first type of behavior to the second type \cite{gopalakrishnan2004,brown2010}.

\begin{figure}[t]
\centering
\includegraphics[width=0.45\columnwidth]{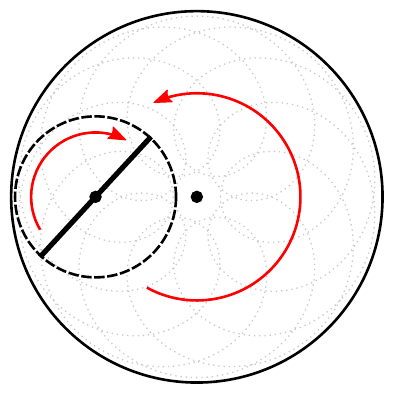}
 \caption{Schematic of a planetary mixer. A blade (bold) rotates inside a bowl, full circle, which counter rotates. Shearing occurs in the gap between the blade and the bowl. }
  \label{fig:Planetary}
\end{figure}

\section*{Conching phenomenology}

We worked with a simplified chocolate formulation of `crumb powder' dispersed in sunflower oil with lecithin \cite{vanDamme2009}. For one experiment, we also added a second surfactant, polyglycerol polyricinoleate (PGPR). Crumb is manufactured by drying a water-based mixture of sucrose crystals, milk and cocoa mass followed by milling \cite{Beckett2003}. To perform a laboratory-scale conche, we used a planetary mixer, \autoref{fig:Planetary}, to prepare \SI{500}{\gram} batches. The total lecithin added was 0.83 wt\%. In the first step, the `dry conche', we mixed the solids with the oil and 0.166 wt\% of lecithin (20\% of the total) in the planetary mixer at $\approx 100$~rpm until the material smears around the bowl right after it has cohered into a single lump around the blade. At $\phi = 0.55$, this took $\approx \SI{40}{\minute}$. Then, for the `wet conche', the remaining lecithin was added and mixed for a further 20 min.

Conched samples were prepared with solid concentrations in the range $0.4 \lesssim \phi \lesssim 0.6$, with $\phi$ calculated using measured densities (see Materials and Methods), so that a weight fraction of 74\% converts to $\phi = 0.55$ (assuming all the fat contained in the crumb melts during conching). In each case, the flow curves of the as-conched sample as well as that of samples successively diluted with pure oil were measured using parallel-plate rheometry (see Materials and Methods).

\begin{figure}[t]
\begin{center}

\includegraphics[width=\columnwidth]{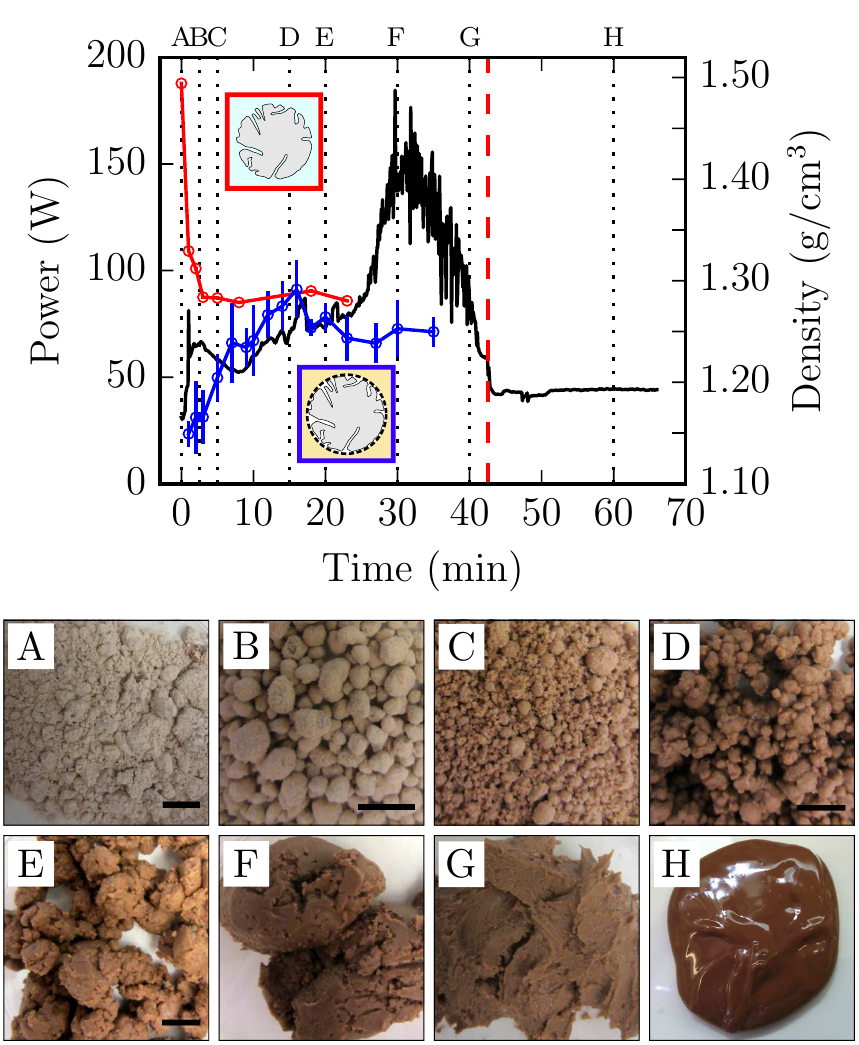}

\caption{\textit{Top} Power consumption ({\protect\tikz[baseline=-0.5ex] \protect\draw [line width=0.5mm, black] (0,0) -- (0.3,0);}), skeletal density ({\protect\tikz[baseline=-0.5ex] \protect\draw [line width=0.5mm, red] (0,0) -- (0.3,0);}) and envelope density ({\protect\tikz[baseline=-0.5ex] \protect\draw [line width=0.5mm, blue] (0,0) -- (0.3,0);}) as a function of mixing time for a typical model chocolate formulation with $\phi_{0} = 0.55$ ($\equiv$~74 wt.\%). Red box: the density of the grey shaded cluster is the skeletal density. Blue box: the average density inside the black dashed circle is the envelope density.  Red dash line denotes time at which second shot of lecithin is added and the transition from dry to wet conche. \textit{Bottom} Visual appearance of samples taken out of the planetary mixer at various stages of the conche.  Letter labels correspond in the upper and lower panels.  Scale bars are \SI{10}{\mm}.  Granule size increases A--E; by F, the granule size has diverged to the size of the system.}
  \label{fig:conche}
\end{center}
\end{figure}

\autoref{fig:conche} (A-H)  show the phenomenology of conching a mixture with solid volume fraction\footnote{That is, the ratio of solid volume to total solid plus liquid volume, discounting any air that may be present; this differs from the granulation literature,  where the air is typically taken into account.}  $\phi_0 = 0.55$ (or 74 wt.\%) to which initially 20\% of the final total of 0.83 wt.\% of lecithin has been added; the accompanying plots show the power consumption of the planetary mixer as well as  measured densities of the sample as conching proceeds. Almost immediately after addition of the sunflower oil to the crumb powder ($t=\SI{0}{\minute}$), all the liquid appeared to have been absorbed. The sample then proceeded to granulate, with the granule size increasing with time. The first granules were visually matt and dry, \autoref{fig:conche}, A-C, and do not stick to each other during mixing.

The skeletal density of a material, \autoref{fig:conche} (red-box inset in plot and Materials and Methods), is the mass of mesoscopic condensed phases (solids and liquids) it contains divided by the volume occupied by these phases, and therefore excludes externally connected air pores. The skeletal density sharply decreased during the first few minutes, converging rapidly to the system average bulk density of the solid and liquid components.  The envelope density,  \autoref{fig:conche} (blue-box inset in plot and Materials and Methods), defined as the mass of a sample divided by its macroscopic volume, including air- and liquid-filled pores, increased over the first \SI{15}{\minute} as the granules compacted, and  converged to the skeletal density. The power consumption increased slowly.

After $\simeq \SI{15}{\minute}$ the granules became visibly moist and coalesced into larger `raspberry-like' structures, \autoref{fig:conche}(D), that somewhat resemble washing powder manufactured by granulation \cite{Salman2006}. The envelope density decreased slightly, presumably due to air incorporation as granules coalesced, and then remained constant: compactification had finished by stages E and F.  The power consumption sharply peaked at $\simeq \SI{30}{\minute}$, when all of the material had formed into a ball adhering to the blade. Thereafter, the power consumption rapidly decreased, and the material became a paste that did not flow easily, \autoref{fig:conche}(G). Soon after this (red dashed line in the plot) the remaining 80\% of lecithin is added, and the `dry conche' transitions to the `wet conche'. The sample rapidly fluidised into a glossy, pourable suspension, \autoref{fig:conche}(H). A sharp decrease in power consumption accompanied this fluidisation.  Note that the consumed power as a function of time is highly reproducible, in both the total power consumed and the time required to reach the peak, provided the same batch of powder is used.

Similar phenomenology is widely reported in wet granulation \cite{Salman2006}, where liquid is gradually added to a dry powder to manufacture granules for applications ranging from agrochemicals to pharmaceuticals. A similar sequence of events to that in \autoref{fig:conche} A-H is seen as the amount of added liquid increases. However, the system would typically become `overwet', i.e.~turn into a flowing suspension, at the peak of the power curve (equivalent to our stage F) \cite{Betz2003}, rather than later (as in our case, at stage H). We note especially that there is a striking visual similarity in the time-lapsed images and power curves for concrete mixing at fixed liquid content \cite{cazacliu2009}. Such similarities across diverse sectors suggest that the incorporation of liquid into powder to form a flowing suspension via various stages of granulation may be underpinned by generic physics, which we seek to uncover through rheology.

\section*{Effect of conching on chocolate rheology} \label{sec:rheology}
\begin{figure}[t]
\centering
  \includegraphics[width=0.8\columnwidth]{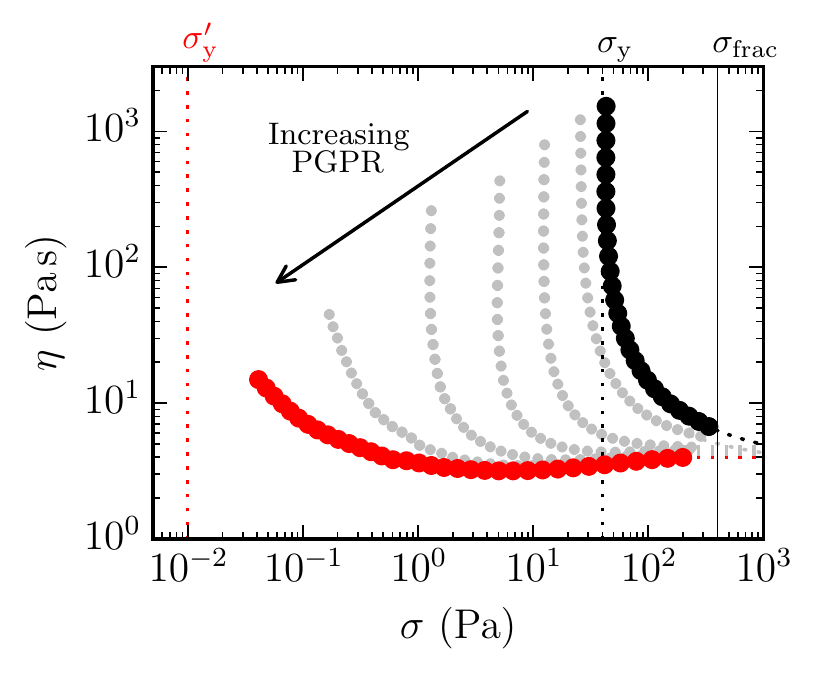}
  \caption{Model chocolate flow curves. ({\color{black}\CIRCLE}): $\sigma^{\star}/\sigma_{\textrm{a}} \ll 1$, crumb conched with \SI{0.83}{\percent} lecithin ($\phi_0 = 0.55$). ({\color{red}\CIRCLE}): $\sigma^{\star}/\sigma_{\textrm{a}} \gg 1$; as before, but with \SI{1.2}{\percent} PGPR.  Intermediate curves for intermediate PGPR contents. Dashed lines guide the eye to high-shear viscosity at $\sigma > \sigma_{\textrm{frac}}$.}
  \label{fig:rheology}
\end{figure}

\autoref{fig:rheology} ({\color{black}\CIRCLE}) shows the flow curve, $\eta(\sigma)$, of a fully-conched crumb mixture with $\phi_0 = 0.54$ (73 wt.\%). Below a yield stress, $\sigma_{\textrm{y}} \approx \SI{40}{\pascal}$, $\eta \to \infty$. Above $\sigma_{\textrm{y}}$ the sample shear thins towards a Newtonian plateau at $\lesssim \SI{5}{\pascal\second}$. However, just before reaching this value, the surface of the sample breaks up and it is no longer contained between the rheometer tools.  This occurs at an approximately $\phi_0$-independent $\sigma_{\rm frac} \approx \SI{400}{\pascal}$, close to the stress, $\sim 0.1\Sigma/a$ ($\approx \SI{300}{\pascal}$ for our system  with $a \approx \SI{10}{\micro\meter}$ and surface tension $\Sigma \approx \SI{30}{\milli\newton\per\meter}$), where particles may be expected to poke out of the free suspension-air interface \cite{brown2014shear}.

Interestingly, if PGPR is added together with the `second shot' of lecithin at the beginning of the wet conche, a different rheology is obtained, \autoref{fig:rheology} ({\color{red}\CIRCLE}): $\sigma_{\textrm{y}}$ is dramatically lowered (here, to $\sigma_{\textrm{y}}^\prime \sim \SI{E-2}{\Pa}$), revealing shear thickening  with an onset stress of $\sigma^\star \gtrsim \SI{2}{\pascal}$. This suggests that in the sample conched with lecithin only, shear thickening is masked \cite{gopalakrishnan2004,brown2010,guy2018constraint} by $\sigma_{\textrm{y}} > \sigma^\star$, but the high-shear viscosity is nevertheless the shear-thickened, frictional-contacts dominated $\eta_2$.

\begin{figure}[t]
\includegraphics[width=0.92\columnwidth]{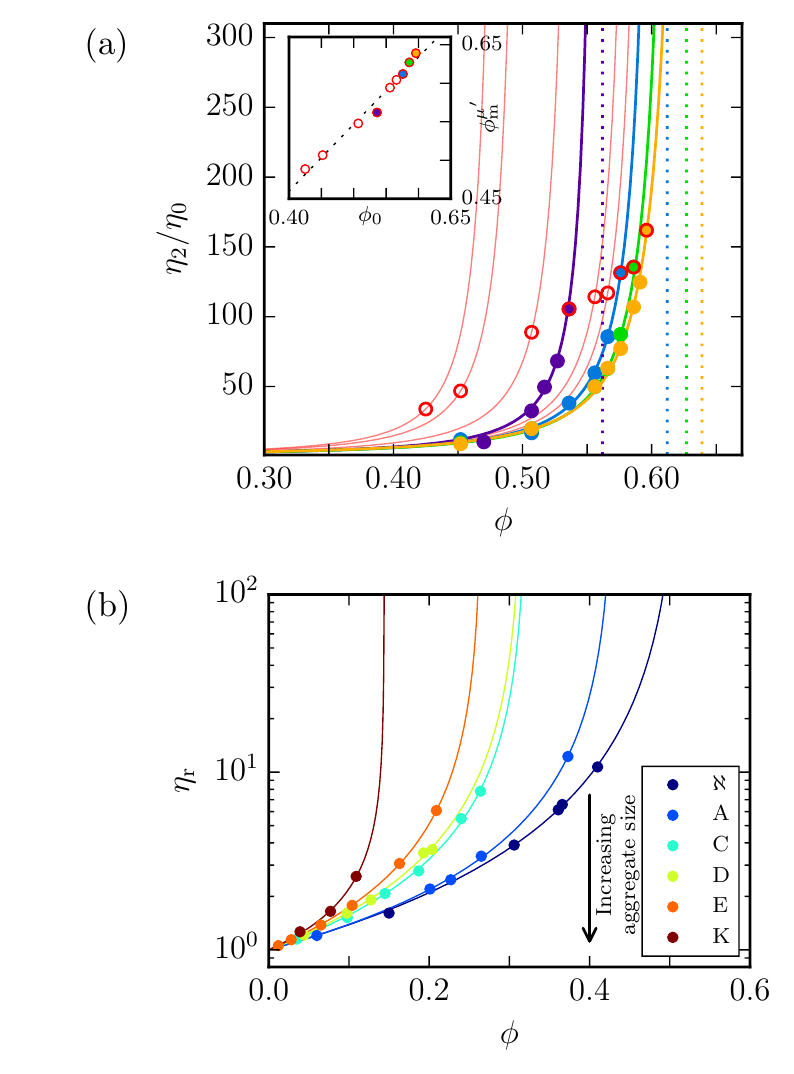}
\caption{(a) Relative high shear viscosity of chocolate suspensions {\it vs} solid volume fraction. ({\color{red}\CircPipe}): $\eta_{\textrm{fc}}^{[\phi_0]}$,   suspensions fully conched in the planetary mixer. \textit{Filled circles}: Diluted fully-conched suspensions: ({\color[RGB]{255,173,0}\CIRCLE}): diluted from $\phi_0 = 0.596$. ({\color[RGB]{0,220,0}\CIRCLE}): diluted from $\phi_0 = 0.586$. ({\color[RGB]{0,120,221}\CIRCLE}): diluted from $\phi_0 = 0.576$. ({\color[RGB]{88,0,159}\CIRCLE}): diluted from $\phi_0 = 0.536$. Matching-color vertical dotted lines: $\phi_{\rm m}$ from fitting \autoref{eq:krieger-dougherty} to these four data sets. Thin red curves: \autoref{eq:krieger-dougherty} with $\lambda = 1,73$ consistent with single open red circle data points. \textit{Inset} Frictional jamming $\phi_{\textrm{m}}$ of chocolate suspensions as a function of the conched volume fraction $\phi_0$. Symbols as in main figure. (b) Replotted data of Lewis and Nielson \cite{lewis1968} for 30-\SI{40}{\micro\meter} glass spheres suspended in Aroclor, with each data set fitted to \autoref{eq:krieger-dougherty}; aggregate size increases from $\aleph$ and then A to K.}
  \label{fig:dilution}
\end{figure}

This high-shear viscosity, $\eta_{\textrm{fc}}^{[\phi_0]}$, of model chocolates fully-conched (fc) with lecithin at nine different solid fractions, $\phi_0$, is plotted in \autoref{fig:dilution}(a) ({\color{red}\CircPipe}). In four cases, we successively diluted the conched samples with sunflower oil and measured the high-shear viscosity along each dilution series, $\eta_{\textrm{fc}}^{[\phi_0]}(\phi)$. Each data set can be fitted to \autoref{eq:krieger-dougherty} with $A=1$, \autoref{fig:dilution}(a) (solid lines), confirming what is already obvious from inspection, viz., that these data sets diverge at different points: {\color[RGB]{255,173,0}\CIRCLE} $\phi_{\textrm{m}} = 0.639$ ($\lambda=1.88$) for $\phi_0 = 0.596$, {\color[RGB]{0,220,0}\CIRCLE} $\phi_{\textrm{m}} = 0.627$ ($\lambda=1.78$) for $\phi_0 = 0.586$, {\color[RGB]{0,120,221}\CIRCLE} $\phi_{\textrm{m}} = 0.612$ ($\lambda=1.72$) for $\phi_0 = 0.576$, {\color[RGB]{88,0,159}\CIRCLE} $\phi_{\textrm{m}} = 0.562$ ($\lambda = 1.53$) for $\phi_0 = 0.536$. In each remaining case, we estimate $\phi_{\textrm{m}}$ without generating a full dilution series at each $\phi_0$ by fitting \autoref{eq:krieger-dougherty} through each of the five $\eta_{\textrm{fc}}^{[\phi_0]}$ data points using the averaged exponent from the four full dilution series $\lambda = \bar \lambda = 1.73$,  \autoref{fig:dilution}(a) (thin red curves).

Plotting all available pairs of $(\phi_0, \phi_{\textrm{m}})$, \autoref{fig:dilution}(a) (inset), confirms that $\phi_{\textrm{m}}$ increases as we conche the mixtures at higher solid fraction $\phi_0$.  That is to say, conching at a higher solid fraction gives rise to a higher jamming volume fraction. There is, however, an upper bound to such $\phi_{\rm m}$ optimisation, which we can estimate by noting that, empirically, $\phi_{\textrm{m}}(\phi_0)$ is approximately linear in our range of $\phi_0$. A linear extrapolation shows that $\phi_{\textrm{m}} = \phi_0$ at $\phi_0^{\textrm{max}} \approx 0.8$. This is likely an overestimate: the approximately linear relation shown in the inset of \autoref{fig:dilution}(a) probably becomes sub-linear and perhaps saturates at higher $\phi_0$. Nevertheless, the existence of some $\phi_0^{\rm max} < 0.8$ beyond which conching will not increase $\phi_{\textrm{m}}$ seems to be a reasonable inference from our data.

\section*{Conching as jamming engineering} \label{sec:conchingJammingEngineering}

\subsection*{Conching reduces aggregate size and increases $\phi_{\textrm{m}}$}

Interestingly, our observations may be interpreted in terms of an `inverse conching' experiment performed some 50 years ago. Lewis and Nielsen \cite{lewis1968} measured the viscosity of 30-\SI{40}{\micro\meter} glass spheres suspended in Aroclor (a viscous Newtonian organic liquid) as a function of volume fraction, \autoref{fig:dilution}(b) (data set $\aleph$), and repeated the measurements with glass beads that have been increasingly aggregated by sintering before dispersal in Aroclor, \autoref{fig:dilution}(b), giving average number of primary particles in an aggregate $N$ of 1.8 (A), 5 (C), 8 (D), 12 (E) and 200-300 (K). Neutral silica in an apolar solvent likely has a vanishingly small $\sigma^\star$, so that Lewis and Nielsen were measuring $\eta_2(\phi)$, the shear-thickened viscosity, which diverges at $\phi_{\textrm{m}}$. Thus, $\phi_{\rm m}$ is clearly lowered by aggregation.

We interpret our data, \autoref{fig:dilution}(a), as `Lewis and Nielsen in reverse'. The primary particles in raw crumb are aggregated in storage due to moisture, etc.~\cite{pietsch2004agglomeration}.  Conching reduces aggregation and therefore increases $\phi_{\textrm{m}}$, with the effect being progressively more marked as the material is being conched at higher $\phi_0$. The latter effect is probably because the same external stress generates higher particle pressure at higher $\phi$ \cite{Boyer2011,Hermes2016}, which breaks up aggregates more effectively.

The linear relation in the inset of \autoref{fig:dilution}(a) extrapolates to a finite intercept at $(0, 0.11\pm0.02)$, suggesting that the viscosity of unconched crumb powder would diverge at $\phi_{\textrm{m}} = 0.11$. This value is perhaps unrealistically low: the real $\phi_{\textrm{m}}(\phi_0)$ dependence probably  becomes sub-linear at low $\phi_0$ and saturates at some value $> 0.11$. Nevertheless, there seems little doubt that unconched, aggregated crumb suspensions jam at volume fractions considerably below those used for real chocolate formulations ($\phi_0 \gtrsim 0.55$).

The change in aggregation during the liquid-powder mixing process is often monitored by laser light scattering. This method would not have separated Lewis and Nielsen's samples $\aleph$, A, C, D and E. Highly accurate data at very low scattering angles are needed to distinguish $N$-mers with small $N$ even when the primary particles are quasi-monodisperse. (See the instructive study of monomers and dimers by Johnson {\it et al.} \cite{Johnson2005}.) Similarly, light scattering will be at best a crude tool for studying conching. In fact, measuring changes in $\phi_{\textrm{m}}$ by rheology is likely  a good, albeit indirect, method to detect changes in low $N$-mer aggregation.

\begin{figure}[t]
\centering
\includegraphics[width=0.9\columnwidth]{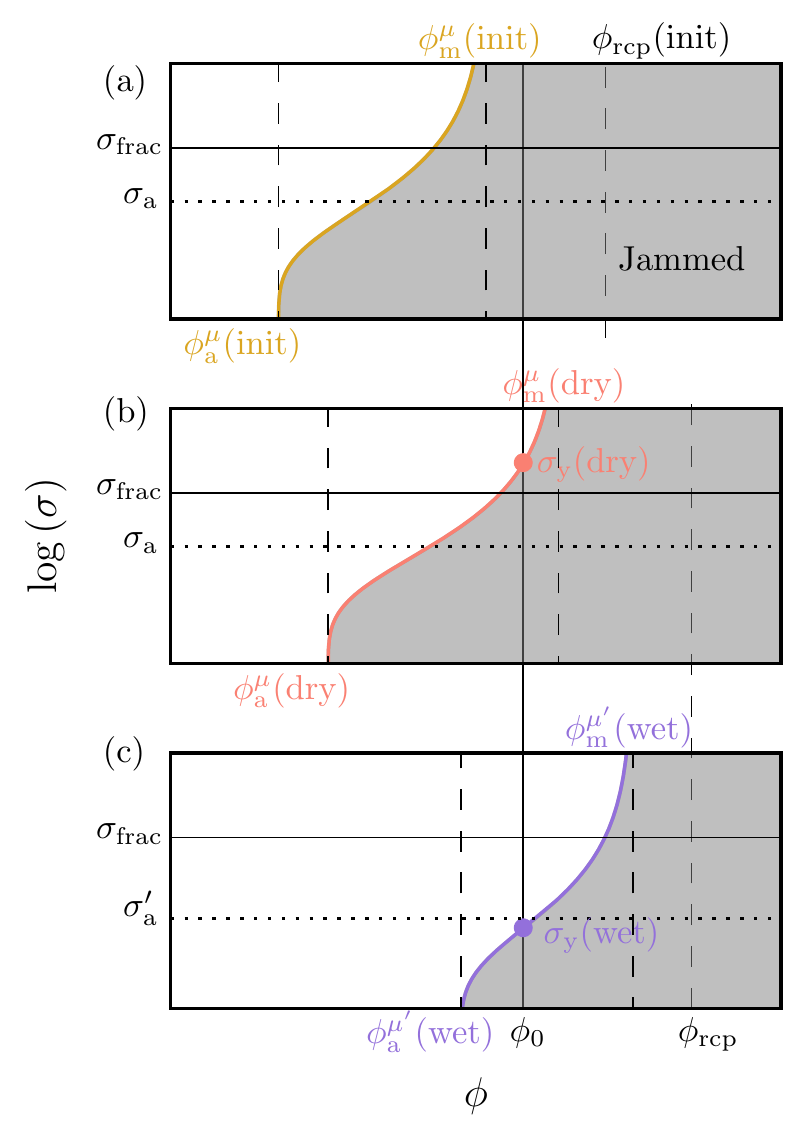}
\caption{The conching process represented as shifts in the jamming state diagram (cf.~\autoref{fig:FrictionDependentPhiJ}(b)). (a) The formulation at $\phi_0$ is considerably more concentrated than the jamming point of the unconched suspension, $\phi_{\textrm{m}}^\mu(\mbox{init})$. (b) It granulates under mechanical agitation, simultaneously breaking up aggregates and increasing both $\phi_{\textrm{m}}^\mu$ and $\phi_{\textrm{a}}^\mu$, thus shifting the jamming boundary to the right. At the end of the dry conche, the suspension could in principle flow, $\phi_0 < \phi_{\textrm{m}}^\mu(\mbox{dry})$, but in fact cannot do so, because $\sigma_{\textrm{frac}} < \sigma_{\textrm{y}}(\mbox{dry})$. (c) Addition of the `second shot' of lecithin reduces the inter-particle friction coefficient to $\mu^\prime < \mu$ and the adhesive interaction to $\sigma_{\textrm{a}}^\prime < \sigma_{\textrm{a}}$, thereby shifting the jamming boundary further to the right and down, dropping $\sigma_{\textrm{y}}(\mbox{wet})$ below $\sigma_{\textrm{frac}}$. Since $\phi_0$ is now considerably below $\phi_{\textrm{m}}^{\mu^\prime}\mbox{(wet)}$, the system is now a flowing suspension. }
  \label{fig:schematic}
\end{figure}

\subsection*{From granules to flowing suspension}

We can now describe the whole conching process in terms of suspension rheology. Consider a crumb-oil-lecithin mixture at $\phi_0 = 0.576$, somewhat higher than in \autoref{fig:conche}.  The jamming point of the initial mixture at time $t = 0$, $\phi_{\textrm{m}}^\mu(\mbox{init})$, is substantially lower (our lower-bound estimate, the $y$-intercept of \autoref{fig:dilution}(a) (inset), being $0.11$). This initial mixture therefore cannot flow homogeneously. In state-diagram terms, the starting system is deep inside the jammed region, \autoref{fig:schematic}(a), where, under mechanical agitation, it granulates \cite{WC2},  \autoref{fig:conche} (B-D). These granules form as a result of there being insufficient liquid to saturate the entire system, and are held together by a combination of surface tension maintaining a jammed particle packing and inter-particle adhesion. The granulation process is controlled by the kinetics of cluster-cluster collisions and the mechanical properties of the clusters \cite{Iveson1998,Iveson2001,Salman2006}. In parallel, aggregates are being broken up, increasing the free volume in the system, so that both $\phi_{\textrm{m}}^\mu$ and $\phi_{\textrm{a}}^\mu$ steadily increase, until $\phi_{\textrm{m}}^\mu$ just exceeds $\phi_0$.

At this point, the system becomes fully saturated and we may expect the system to turn into a flowable suspension, albeit with a very high viscosity, since there is no longer a shear jammed state for surface tension to maintain. This is indeed what happens in many systems: the power is observed to peak just as the material becomes (in granulation jargon) `overwet' \cite{Betz2003}, i.e.~a flowing suspension with a shiny surface. In our case, at the power peak (stage F in \autoref{fig:conche}), the suspension still does not flow easily and appears visually matt. This is probably because the sample fractures before it can yield to flow homogeneously, i.e., $\sigma_{\textrm{y}} > \sigma_{\textrm{frac}}$ (cf.~the earlier discussion of $\sigma_{\textrm{frac}}$ associated with \autoref{fig:rheology}).

Further conching continues to increase $\phi_{\textrm{m}}$ until at the end of the dry conche, the yield stress, $\sigma_{\textrm{y}}(\mbox{dry})$, only just exceeds the fracture stress, $\sigma_{\textrm{frac}}$, \autoref{fig:schematic}(b). Here, the addition of the `second shot' of lecithin has a dramatic effect. We suggest that this is because the additional lecithin lowers $\mu$ and $\sigma_{\textrm{a}}$ to $\mu^\prime$ and $\sigma_{\textrm{a}}^\prime$. The jamming boundary abruptly shifts to the right and drops down, \autoref{fig:schematic}(c). The resulting dramatic lowering of the yield stress to $\sigma_{\textrm{y}}(\mbox{wet}) < \sigma_{\textrm{frac}}$ in a system where, now, $\phi_0$ is considerably below $\phi_{\textrm{m}}(\mbox{wet})$, immediately produces a flowing suspension, liquid chocolate.

\subsection*{Lecithin as a lubricant}

\begin{figure}[t]
\centering
\includegraphics[width=0.7\columnwidth]{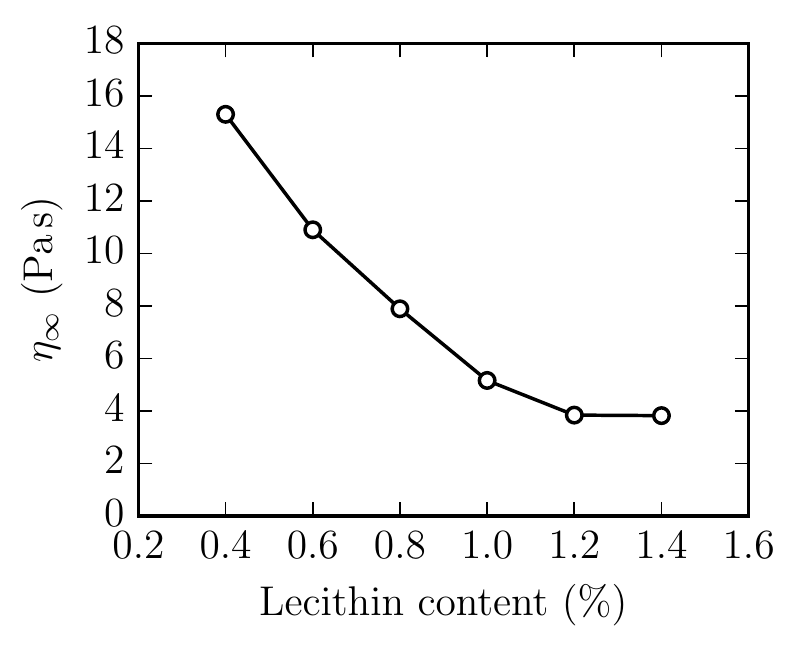}
\caption{High-shear viscosity of model chocolate  at $\phi_0 = 0.557$ dry-conched without lecithin as a function of subsequently-added lecithin. }
\label{fig:RoleOfLecithin}
\end{figure}

We have suggested that the lecithin added in the `second shot' lowers $\mu$ and therefore increases both $\phi_{\textrm{m}}^\mu$ and $\phi_{\textrm{a}}^\mu$ by releasing constraints on the system \cite{guy2018constraint}.   To provide direct experimental evidence for this role, we prepared a dry conche with the first shot of lecithin omitted, which again produced a non-flowing paste. Various amounts of lecithin was mixed into aliquots of this paste, which liquified. The high-shear viscosity, $\eta_2$, of the resulting suspensions decreases with lecithin concentration, \autoref{fig:RoleOfLecithin}. To check that this was not due to the oils in lecithin lowering the sample volume fraction, we repeated the experiment added an equivalent volume of oil corresponding to the maximum lecithin concentration (1.4\%). This failed to liquify the paste. We may therefore conclude that lecithin causes this effect by lowering $\mu$ and so increasing the jamming point, $\phi_{\textrm{J}} = \phi_{\textrm{m}}^\mu$, \autoref{fig:FrictionDependentPhiJ} (inset).

\section*{Summary and conclusions}

Creating flowable solid-in-liquid dispersions of maximal solid content is a generic goal across many industrial sectors. We have studied one such process in detail, the conching of crumb powder and sunflower oil into a flowing model chocolate. We interpreted our observations and measurements using existing knowledge from the granulation literature as well as an emerging new understanding of shear thickening and jamming in granular dispersions. The resulting picture is summarised in \autoref{fig:schematic}. The essential idea is that conching, and more generally wet milling, is about `jamming engineering' -- manipulating $\phi_{\textrm{m}}^\mu$ and $\phi_{\textrm{a}}^\mu$ by changing the state of aggregation, and `tuning' the inter-particle friction coefficient $\mu$ and the strength of inter-particle adhesion, $\sigma_{\textrm{a}}$. Importantly, many additives ostensibly acting as dispersants to reduce inter-particle attraction and so lower $\sigma_{\rm a}$ may, in fact, function primarily as lubrications to lower $\mu$ and so increase $\phi_{\rm m}$. Our scheme, \autoref{fig:schematic}, with appropriate shifts in $\sigma_{\textrm{frac}}$, can be used to understand liquid incorporation into powders in many different specific applications.

Our proposed picture for conching/wet milling poses many questions. For example, the rheology of a suspension at the end of dry conche, in which flow is in principle possible ($\phi_0 < \phi_{\textrm{m}}$) but in practice ruled out by surface fracture occurring before bulk yielding, has not yet been studied in any detail. Neither is the role of changing $\phi_{\textrm{m}}$ during the granulation process understood. Our results therefore constitute only a first step towards a unified description of liquid incorporation, wet milling and granulation.




\matmethods{
\label{sec:material}

Our crumb powder (supplied by Mars Chocolate UK) consists mostly of faceted particles with mean radius $a \approx \SI{10}{\micro\meter}$ (polydispersity $\gtrsim \SI{150}{\percent}$) according to laser diffraction (Beckman-Coulter LS-13 320). It has a density \SI{1.453}{\gram\per\cubic\centi\meter} and a specific (BET) surface area \SI{2.02}{\square\meter\per\gram} (data provided by Mars Chocolate UK). We used sunflower as purchased (Flora) and soy lecithin as supplied (by Mars Chocolate UK). The latter consists of a mixture of phospholipids ($\approx \SI{60}{\percent}$) with some residual soya oil. Our sunflower oil was Newtonian at \SI{20}{\celsius}, with viscosity $\eta_0 = \SI{54}{\milli\pascal\second}$. Its density was measured by an Anton Paar DMA density meter to be \SI{0.917}{\gram\per\cubic\centi\meter}. Using sunflower oil rather than cocoa butter obviates the need for heating during rheological measurements. The rheology of the chocolate suspension obtained from conching our mixture resembles that of fresh liquid chocolate made using cocoa butter \cite{vanDamme2009}. Our PGPR was also supplied by Mars Chocolate UK.

Conching was performed using a Kenwood kMix planetary mixer with a K-blade attachment, adding the lecithin in two successive batches as described in the main text. We measured the skeletal density at various stages of conching by performing helium pycnometry (Quantachrome Ultrapyc). The envelope density was measured using a 2.00 $\pm$ \SI{0.02}{\ml} Pyrex micro-volumetric flask with sunflower oil as the liquid phase.

Rheometric measurements were performed using a stress-controlled rheometer (TA Instruments, DHR-2) in a cross-hatched plate-plate geometry (diameter 40 mm, $1\times1\times\SI{0.5}{\milli\meter}$ serrated grid of truncated pyramids) to minimise wall slip, at a gap height of \SI{1}{\mm} and a temperature of \SI{20}{\celsius}.

All data plotted in this work can be downloaded from \href{https://doi.org/10.7488/ds/2515}{https://doi.org/10.7488/ds/2515}.
}

\showmatmethods 

\acknow{We thank Mars Chocolate UK Ltd.~for initiating and funding part of this work. Other funding came from the EPSRC (EP/J007404/1 and EP/N025318/1). Research at NYU was supported partially by the MRSEC Program of the National Science Foundation under Award Number DMR-1420073.  We thank Ben Guy, John Royer and Jin Sun (Edinburgh) and Will Taylor (Mars Chocolate) for illuminating discussions.}

\showacknow 

\pnasbreak


\bibliography{Bibliography_conching_final}

\begin{thebibliography}{10}

\bibitem{Kaye1997}
Kaye B (1997) {\em Powder Mixing}.
\newblock (Springer Netherlands).

\bibitem{Green2004}
Carneim TJ, Green DJ (2004) Mechanical properties of dry-pressed alumina green
  bodies.
\newblock {\em J. Am. Ceramic Soc.} 84:1405--1410.

\bibitem{Tao2016}
Tao R, Tang H, Tawhid-Al-Islam K, Du E, Kim J (2016) Electrorheology leads to
  healthier and tastier chocolate.
\newblock {\em Proc. Natl. Acad. Sci. (USA)} 113:7399--7402.

\bibitem{Gutierrez2017}
Guti{\'e}rrez TJ (2017) State-of-the-art chocolate manufacture: A review.
\newblock {\em Comp. Rev. Food Safety} 16:1313--1344.

\bibitem{Seto2013}
Seto R, Mari R, Morris JF, Denn MM (2013) Discontinuous shear thickening of
  frictional hard-sphere suspensions.
\newblock {\em Phys. Rev. Lett.} 111:218301.

\bibitem{WC1}
Wyart M, Cates ME (2014) Discontinuous shear thickening without inertia in
  dense non-{B}rownian suspensions.
\newblock {\em Phys. Rev. Lett.} 112:098302.

\bibitem{WC2}
Cates ME, Wyart M (2014) Granulation and bistability in non-{B}rownian
  suspensions.
\newblock {\em Rheol. Acta} 53:755--764.

\bibitem{mari2014shear}
Mari R, Seto R, Morris JF, Denn MM (2014) Shear thickening, frictionless and
  frictional rheologies in non-{B}rownian suspensions.
\newblock {\em J. Rheol.} 58(6):1693--1724.

\bibitem{Guy2015}
Guy BM, Hermes M, Poon WCK (2015) Towards a unified description of the rheology
  of hard-particle suspensions.
\newblock {\em Phys. Rev. Lett.} 115:088304.

\bibitem{Lin2015}
Lin NY et~al. (2015) Hydrodynamic and contact contributions to continuous shear
  thickening in colloidal suspensions.
\newblock {\em Phys. Rev. Lett.} 115:228304.

\bibitem{Royer2016}
Royer JR, Blair DL, Hudson SD (2016) Rheological signature of frictional
  interactions in shear thickening suspensions.
\newblock {\em Phys. Rev. Lett.} 116:188301.

\bibitem{Hermes2016}
Hermes M et~al. (2016) Unsteady flow and particle migration in dense,
  non-{B}rownian suspensions.
\newblock {\em J. Rheol.} 60:905--916.

\bibitem{Colin2017}
Comtet J et~al. (2017) Pairwise frictional profile between particles determines
  discontinuous shear thickening transition in non-colloidal suspensions.
\newblock {\em Nat. Commun.} 8:15633.

\bibitem{clavaud2017revealing}
Clavaud C, B{\'e}rut A, Metzger B, Forterre Y (2017) Revealing the frictional
  transition in shear-thickening suspensions.
\newblock {\em Proceedings of the National Academy of Sciences} p. 201703926.

\bibitem{Maron1956}
Maron SH, Pierce PE (1956) Application of ree-eyring generalized flow theory to
  suspensions of spherical particles.
\newblock {\em J. Colloid Sci.} 11:80 -- 95.

\bibitem{KD1959}
Krieger IM, Dougherty TJ (1959) A mechanism for non‐newtonian flow in
  suspensions of rigid spheres.
\newblock {\em Trans. Soc. Rheol.} 3:137--152.

\bibitem{silbert2010jamming}
Silbert LE (2010) Jamming of frictional spheres and random loose packing.
\newblock {\em Soft Matter} 6(13):2918--2924.

\bibitem{Hodgson2015}
Hodgson DJM, Hermes M, Poon WCK (2015) Jamming and the onset of granulation in
  a model particle system.
\newblock {\em arXiv:1507.08098}.

\bibitem{guy2018constraint}
Guy BM, Richards JA, Hodgson DJM, Blanco E, Poon WCK (2018) Constraint-based
  approach to granular dispersion rheology.
\newblock {\em Phys. Rev. Lett.} 121(12):128001.

\bibitem{liu2017}
Liu W, Jin Y, Chen S, Makse HA, Li S (2017) Equation of state for random sphere
  packings with arbitrary adhesion and friction.
\newblock {\em Soft Matter} 13(2):421--427.

\bibitem{gopalakrishnan2004}
Gopalakrishnan V, Zukoski C (2004) Effect of attractions on shear thickening in
  dense suspensions.
\newblock {\em Journal of Rheology} 48(6):1321--1344.

\bibitem{brown2010}
Brown E et~al. (2010) Generality of shear thickening in dense suspensions.
\newblock {\em Nature Mat.} 9(3):220.

\bibitem{vanDamme2009}
Taylor JE, {Van Damme} I, Johns ML, Routh AF, Wilson DI (2009) Shear rheology
  of molten crumb chocolate.
\newblock {\em J. Food Sci.} 74:55--61.

\bibitem{Beckett2003}
Beckett ST (2003) Is the taste of british milk chocolate different.
\newblock {\em Int. J. Dairy Tech.} 56:139--142.

\bibitem{Salman2006}
Salman AD, Hounslow M, Seville JP, eds. (2006) {\em Granulation}, Handbook of
  Powder Technology.
\newblock (Elsevier) Vol.{}~11.

\bibitem{Betz2003}
Betz G, B{\"u}rgin PJ, Leuenberger H (2003) Power consumption profile analysis
  and tensile strength measurements during moist agglomeration.
\newblock {\em Int. J. Pharm.} 252(1):11 -- 25.

\bibitem{cazacliu2009}
Cazacliu B, Roquet N (2009) Concrete mixing kinetics by means of power
  measurement.
\newblock {\em Cement Concrete Res.} 39:182--194.

\bibitem{brown2014shear}
Brown E, Jaeger HM (2014) Shear thickening in concentrated suspensions:
  phenomenology, mechanisms and relations to jamming.
\newblock {\em Rep. Prog. Phys.} 77(4):046602.

\bibitem{lewis1968}
Lewis TB, Nielsen LE (1968) Viscosity of dispersed and aggregated suspensions
  of spheres.
\newblock {\em Trans. Soc. Rheol.} 12:421--443.

\bibitem{pietsch2004agglomeration}
Pietsch W (2004) {\em Agglomeration in Industry, 2 Volume Set: Occurrence and
  Applications}.
\newblock (John Wiley \& Sons).

\bibitem{Boyer2011}
Boyer F, Guazzelli {\'E}, Pouliquen O (2011) Unifying suspension and granular
  rheology.
\newblock {\em Phys. Rev. Lett.} 107:188301.

\bibitem{Johnson2005}
Johnson PM, van Kats CM, van Blaaderen A (2005) Synthesis of colloidal silica
  dumbbells.
\newblock {\em Langmuir} 21(24):11510--11517.

\bibitem{Iveson1998}
Iveson SM, Litster JD (1998) Growth regime map for liquid-bound granules.
\newblock {\em AIChE J.} 44:1510--1518.

\bibitem{Iveson2001}
Iveson S et~al. (2001) Growth regime map for liquid-bound granules: Further
  development and experimental validation.
\newblock {\em Powder Technol.} 117:83--97.

\end{thebibliography}

\end{document}